\documentclass[aps,prl,twocolumn,groupedaddress,showpacs]{revtex4-1}
\usepackage{graphicx}
\usepackage{dcolumn}
\usepackage{bm}
\usepackage{amsmath}
\usepackage{amssymb}
\usepackage{color}
\usepackage{subfigure}
\usepackage{multirow}

\begin{document}

\title{Experimental observation of topological transitions in interacting multi-spin systems}

\author{Zhihuang Luo $^1$}
\author{Chao Lei $^1$}
\author{Jun Li $^2$}
\author{Xinfang Nie $^1$}
\author{Zhaokai Li $^1$}
\author{Xinhua Peng $^{1,3}$}
\email{xhpeng@ustc.edu.cn}
\author{Jiangfeng Du $^{1,3}$}
\email{djf@ustc.edu.cn}
\affiliation{$^1$Hefei National Laboratory for Physical Sciences at Microscale and Department of Modern Physics, University of Science and Technology of China, Hefei, Anhui 230026, China}
\affiliation{$^{2}$Beijing Computational Science Research Center, Beijing, 100094, China}
\affiliation{$^{3}$Synergetic Innovation Center of Quantum Information $\&$ Quantum Physics, University of Science and Technology of China, Hefei, Anhui 230026, China}

%\date{\today}

\begin{abstract}
Topologically ordered phase has emerged as one of most exciting concepts that not only broadens our understanding of phases of matter, but also has been found to have potential application in fault-tolerant quantum computation. The direct measurement of topological properties, however, is still a challenge especially in interacting quantum system. Here we realize one-dimensional Heisenberg spin chains using nuclear magnetic resonance simulators and observe the interaction-induced topological transitions, where Berry curvature in the parameter space of Hamiltonian is probed by means of dynamical response and then the first Chern number is extracted by integrating the curvature over the closed surface. The utilized experimental method provides a powerful means to explore topological phenomena in quantum systems with many-body interactions.
\end{abstract}


\maketitle

Since the first observation of topologically ordered phases in quantum Hall effect in 1980s \cite{Klizing1980,Tsui1982}, there is growing interest in studying the topology of quantum systems such as topological insulators \cite{Zhang2006,Kane2007,Zhang2007,Hsieh2008,Levin2009} and spin liquids \cite{Laughlin1987,Wen1991,Moessner2001,Balents2010}. Meanwhile, the great efforts in fault-tolerant quantum computation are being made on the basis of the existence of topologically ordered phases \cite{Kitaev2003,Nayak2008,Stern2013}. The different topological phases and their topological transitions are characterized by robust topological invariants in physics, most of which arise as integrals of some geometric quantity. For example, the first Chern number \cite{Chern1946}, the integral of Berry curvature over the closed surface of parameter space of Hamiltonian, is a well-defined topological invariant. It is closely related to Berry phase \cite{Berry1984,Niu2010}. As emerged in quantum Hall physics, the filling factor known as first Chern number in mathematics is used to distinguish different quantum Hall states \cite{Hatsugai1993,Avron2003}. When the jumps of their values happen, it undergos quantum Hall transitions. Topological invariant reveals the global properties of topological phases and remains unchanged under small perturbations.

However, it is still an experimental challenge to directly probe the topological properties especially in interacting quantum systems. Usually, the previous measurement of Berry phase relies on the interference experiment \cite{Leek2007,Peng2010,Berger2012,Berger2013}, but this method is limited to systems of weakly interacting quasi-particles. Beyond this limit, Gritsev \textit{et al}. \cite{Gritsev2012} proposed an alternative method to directly measure the Berry curvature via the nonadiabatic response on physical observables to the rate of change of an external parameter. Direct measurement of Berry curvature provides a powerful and generalizable means to explore topological properties in any quantum systems where the Hamiltonian can be written in terms of a set of parameter. Based on this method, some experimental observations of topological transition have been demonstrated in small (one- or two-qubit) superconducting systems \cite{Schroer2014,Roushan2014}, and also an experimental scheme was proposed to simulate dynamical quantum Hall effect in Heisenberg spin chain with interacting superconducting qubits \cite{Zhu2015}.

In this Letter, we use several nuclear spins to simulate one-dimensional Heisenberg spin chain that was proposed in Ref.\cite{Gritsev2012} to have the quantization of first Chern number in dynamical response. The emergent different quantized plateaus are related to different topological phases and the interaction-induced topological transitions are observed in nuclear magnetic resonance (NMR) systems. In the experiments, we measure the total magnetization vectors perpendicular to quench velocity by decoupling. The Berry curvatures in parameter space of Hamiltonian are extracted via the linear response and then the first Chern numbers are obtained by integrating the closed surface. From the resulting Chern number, we can visualize the geometric structure of Hamiltonian. The precise quantization of first Chern number may be applied to parameter estimation of Hamiltonian.
The full controllability of NMR will make it possible to experimentally investigate many-body phenomena such as the even-odd effect of Heisenberg chains \cite{Politi2009,Oh2010}.

The first Chern number is defined as the integral of Berry curvature $\mathcal{F}_{\mu\nu}$ over a closed manifold $\mathcal{S}$ in the parameter space $\vec{R}$ of Hamiltonian as \cite{Niu2010,Gritsev2012}
\begin{equation}\label{ch1}
    \mathcal{C}h_1=\frac{1}{2\pi}\oint_{\mathcal{S}}dS_{\mu\nu}\mathcal{F}_{\mu\nu},
\end{equation}
where $\mathcal{F}_{\mu\nu}=\partial_{\mu}\mathcal{A}_{\nu}-\partial_{\nu}\mathcal{A}_{\mu}$, and $\mathcal{A}_{\mu}=i\langle\psi_0|\partial_{\mu}|\psi_0\rangle$. Here we use the shorten notations, i.e., $\partial_{\mu}\equiv\partial_{R_{\mu}}$ and consider the ground state $|\psi_0\rangle$. Berry connection $\mathcal{A}_{\mu}$ and Berry curvature $\mathcal{F}_{\mu\nu}$ can be viewed as a local gauge potential and gauge field, respectively. In analogy to electrodynamics, the local gauge-dependent Berry connection can never be physically observable. Whereas Berry curvature is gauge-invariant and may be related to physical observable that manifests the local geometric property of the ground state in the parameter space. While the first Chern number shows the global topological property of the ground state manifold as a whole. In fact, $\mathcal{C}h_1$ exactly counts the number of degenerate ground states enclosed by parameter space $\mathcal{S}$. To see this point more intuitively, we substitute $\mathcal{A}_{\mu}$ into $\mathcal{F}_{\mu\nu}$ and rewrite the Berry curvature as follows \cite{Niu2010},
\begin{equation}\label{bcurv}
    \mathcal{F}_{\mu\nu}=i\sum_{n\neq 0}\frac{\langle\psi_0|\partial_{\mu}\hat{\mathcal{H}}|\psi_n\rangle\langle\psi_n|
    \partial_{\nu}\hat{\mathcal{H}}|\psi_0\rangle-(\nu\leftrightarrow\mu)}{(\varepsilon_n-\varepsilon_0)^2}.
\end{equation}
Here $\varepsilon_n$ and $|\psi_n\rangle$ are the $n_{th}$ eigenvalue and its corresponding eigenstate of Hamiltonian $\hat{\mathcal{H}}$, respectively. From Eq. (\ref{bcurv}), it clearly shows that degeneracies (i.e., $\varepsilon_n=\varepsilon_0$) are some singular points that will contribute nonzero terms to the integral of $\mathcal{F}_{\mu\nu}$, that is, Eq. (\ref{ch1}). These degeneracy points act as the sources of $\mathcal{C}h_1$ and are analogous to magnetic monopoles in parameter space. The first Chern number is essential for understanding of the quantized effect. It can be used as the nontrivial order parameter to characterize different topological phases and their transitions \cite{Niu2010}.

The usual interference experiments for measuring Berry phase \cite{Leek2007,Peng2010,Berger2012,Berger2013} do not readily generalize to more complicated interacting quantum systems. We follow an alternative approach as proposed in Ref. \cite{Gritsev2012}. It states that Berry curvature can be extracted from the linear response of generalized force $\mathcal{M}_{\mu}$ along the $\mu$-direction, i.e.,
\begin{equation}\label{Mu}
    \mathcal{M}_{\mu}=\text{const}+\mathcal{F}_{\mu\nu}v_{\nu}+\mathcal{O}(v^2),
\end{equation}
where $\mathcal{M}_{\mu}=-\langle\psi_0(t_f)|\partial_{\mu}\hat{\mathcal{H}}|\psi_0(t_f)\rangle$, and $v_{\nu}$ is the quench velocity. To neglect the nonlinear term, the chosen $v_{\nu}$ should be small enough or quasiadiabatic.

\begin{figure}
  \includegraphics[width=8.5cm]{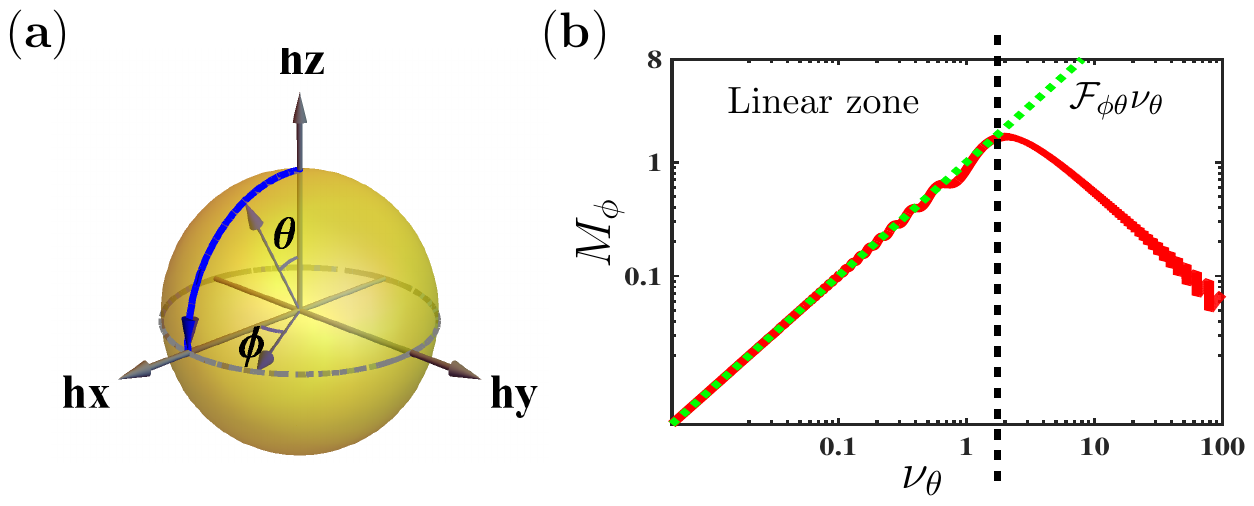}\\
  \caption{(Color online) (a) The quasiadiabatic evolution path (blue) in the spherical parameter space of external magnetic field $\vec{h}$ ($|\vec{h}|=1$). (b) The magnetization as the function of $v_{\theta}$ for two-qubit case. In the linear zone (i.e., at the left of dashed black line), $M_{\phi}=\mathcal{F}_{\phi\theta}v_{\theta}$ and $v_{\theta}^{\text{max}}=1.53$. The high order terms of $v_{\theta}$ dominate at the right of dashed black line. }\label{fig:1}
\end{figure}

The one-dimensional Heisenberg spin chain can be taken as the example to demonstrate the above idea \cite{Gritsev2012,Zhu2015}, whose Hamiltonian in an external magnetic field $\vec{h}$ is described by
\begin{equation}\label{Ham}
    \hat{\mathcal{H}}=-\sum_{j=1}^N\vec{h}\cdot\vec{\sigma}-J\sum_{j=1}^{N-1}\vec{\sigma}_j\cdot\vec{\sigma}_{j+1},
\end{equation}
where $\vec{\sigma}\equiv(\hat{\sigma}_x,\hat{\sigma}_y,\hat{\sigma}_z)$ stands for Pauli matrices, and $J$ is the isotropic coupling interaction strength between the nearest-neighbor spins. In order to measure Berry curvature, let the system start with the initial ground state at the north pole of spherical parameter space of external magnetic field $\vec{h}$ (here we fixed $|\vec{h}|=1$), and then undergo a quasiadiabatic evolution along the blue path, as illustrated in Fig. \ref{fig:1}(a). The path is determined by fixing $\phi=0$ and varing $\theta(t)=v_{\theta}^2t^2/2\pi$ from $t=0$ to $t=\pi/v_{\theta}$ \cite{Gritsev2012}. This choice guarantees that the angular velocity is turned on smoothly and the system is not excited at the beginning of the evolution. The generalized force at $t=\pi/v_{\theta}$ is easily derived as
\begin{equation}\label{}
      \mathcal{M}_{\phi}=-\langle\partial_{\phi}\hat{\mathcal{H}}\rangle|_{\phi=0,t=\pi/v_{\theta}}=\sum_{j=1}^N\langle\hat{\sigma}_y^j\rangle.
\end{equation}
That is equivalent to total magnetization along $y$ direction. Figure \ref{fig:1}(b) shows the limitation of $v_{\theta}$  calculated in two-qubit case, with similar results in three- and four-qubit cases. In the linear zone of $v_{\theta}\leq 1.53$, $M_y\propto v_{\theta}$ and the linear response approximation works well. Then Berry curvature can be obtained from Eq. (\ref{Mu}). By integrating $\mathcal{F}_{\phi\theta}$ over the sphere of $|\vec{h}|=1$, we get the first Chern number. Due to the rotational invariance of the interaction in Eq. (\ref{Ham}), the integration will be simplified into the multiplication of $\mathcal{F}_{\phi\theta}$ by the spherical area $4\pi$ \cite{Gritsev2012}. So we have $\mathcal{C}h_1=2\mathcal{F}_{\phi\theta}$ from Eq. (\ref{ch1}).

\begin{figure}
  \includegraphics[width=8.5cm]{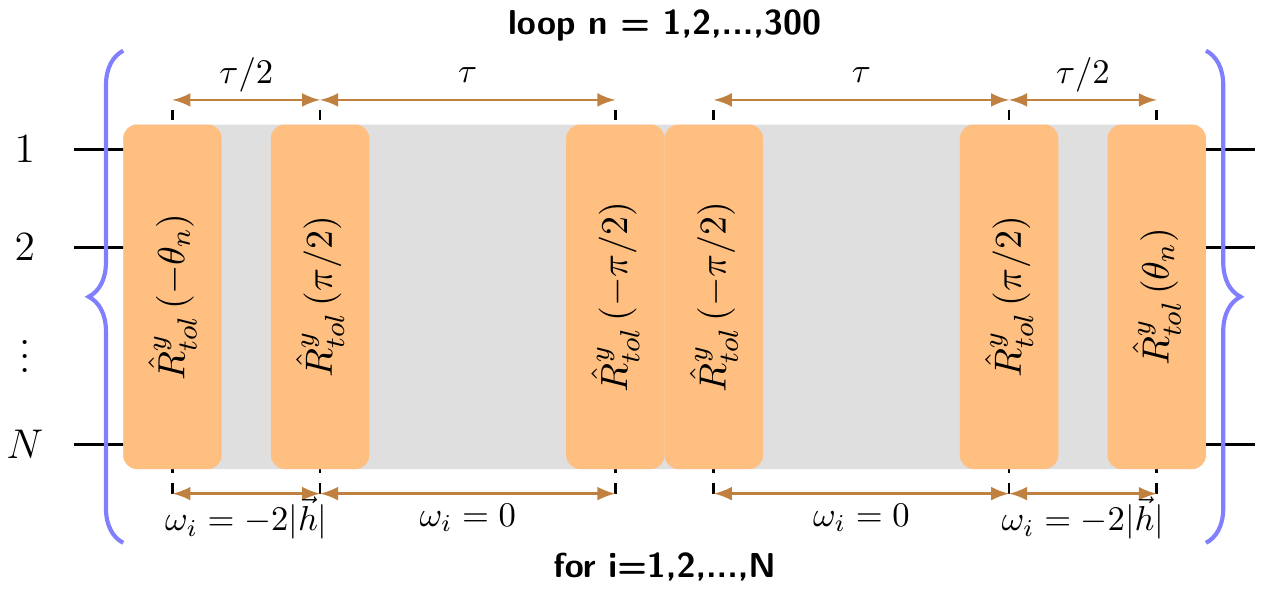}\\
  \caption{(Color online) The pulse sequence for simulating the one-dimensional Heisenberg model of Eq. (\ref{Ham}). The gray regions represent the free evolution of isotropic z-coupling interaction, i.e., $\hat{\mathcal{H}}_{zz}=-J\sum_{j=1}^{N-1}\hat{\sigma}_j^z\hat{\sigma}_{j+1}^z$. They are in different rotating frames by means of setting the off-resonace frequencies of RF pulses acting on $N$ different nuclei, which satisfy $\omega_i=-2|\vec{h}|$ during the $\tau/2$ and $\omega_i=0$ during the $\tau$ for $i=1,2,\cdots,N$, respectively. $\hat{R}_{\text{tol}}^{\alpha}(\theta_n)=\prod_{j=1}^Ne^{-i\theta_n\hat{\sigma}_j^{\alpha}/2} (\alpha=x,y)$ and $\theta_n=v_{\theta}^2t_n^2/2\pi$ for $n=1,2,\cdots,300$. The loop is used to approximate the quasiadiabatic evolution path.}\label{fig:2}
\end{figure}

The Heisenberg model can be effectively simulated by NMR system, whose natural Hamiltonian in the rotating frame is
\begin{equation}\label{HamNMR}
   \hat{\mathcal{H}}_{\text{NMR}}=\sum_{i=1}^{N}\frac{\omega _{i}}{2}\hat{\sigma}^{z}_{i}+\sum_{i<j,=1}^{N}\frac{\pi J_{ij}}{2}\hat{\sigma} ^{z}_{i}\hat{\sigma} ^{z}_{j},
\end{equation}
where $\omega _{i}$ represents the chemical shift of spin $i$ and $J_{ij}$ the coupling constant between spin $i$ and spin $j$. Compared to Eq. (\ref{Ham}), the Hamiltonian of NMR system has the similar form. It is suitable for a NMR system to simulate the Heisenberg model \cite{Peng2005}. According to average Hamiltonian theory \cite{Haeberlen1968}, we can design NMR pulse sequences to effectively create the desired Hamiltonian. The pulse sequences are shown in Fig. \ref{fig:2}(a). Using Trotter approximation, we have in a short period of $\tau$,
\begin{align}
    e^{-i\hat{\mathcal{H}}\tau}=&\hat{R}_{\text{tol}}^y(\theta_n)e^{-i(\hat{\mathcal{H}}_z+\hat{\mathcal{H}}_{zz})\tau/2}
    e^{-i(\hat{\mathcal{H}}_{xx}+\hat{\mathcal{H}}_{yy})\tau} \\ \nonumber
    &\cdot e^{-i(\hat{\mathcal{H}}_z+\hat{\mathcal{H}}_{zz})\tau/2}\hat{R}_{\text{tol}}^y(-\theta_n)+O(\tau^3),
\end{align}
where $\hat{\mathcal{H}}_z=-\sum_{j=1}^N|\vec{h}|\hat{\sigma}_j^z$ and $\hat{\mathcal{H}}_{\alpha\alpha}=-J\sum_{j=1}^{N-1}\hat{\sigma}_j^{\alpha}\hat{\sigma}_{j+1}^{\alpha}$ $(\alpha=x,y,z)$. During $\tau/2$, i.e., at the front and back gray regions in Fig. \ref{fig:2}, the off-resonance frequencies of radio-frequency (RF) pulses acting on $N$ different nuclei are set to satisfy $\omega_i=-2|\vec{h}|$ for $i=1,2,\cdots,N$. In this rotating frame, $\hat{\mathcal{H}}_z$ is turned on during the free evolution of $\hat{\mathcal{H}}_{zz}$. When $\omega_i=0$ for $i=1,2,\cdots,N$, $\hat{\mathcal{H}}_z$ is turned off. $\hat{\mathcal{H}}_{xx}/\hat{\mathcal{H}}_{yy}$ can be readily realized by rotating $\hat{\mathcal{H}}_{zz}$. To implement $\hat{\mathcal{H}}_{zz}$, it only requires to refocus some unwanted interactions in natural Hamiltonian (\ref{HamNMR}) and tune $J_{ij}$s into an isotropic coupling constant. This task can be implemented only using some refocusing $\pi$ pulses. As shown in Appendix B, two examples of $N=3$ and $N=4$ were given. For $N=2$, $\hat{\mathcal{H}}_{zz}$ is the natural interaction of NMR system. We will observe topological transitions in two-, three- and four-spin interacting systems as follows.

In the experiments, we need to measure the total magnetization. However, it will become a challenge when considering the multi-spin systems with interaction. Within the linear zone of $v_{\theta}\leq 1.53$, the generated $M_y$ under the quasiadiabatic evolution is small and will tend to zero in the adiabatic limit, i.e., $v_{\theta}\rightarrow 0$. In addition, one can only measure the magnetization of each spin and its NMR signal will further split into $2^{N-1}$ peaks induced by the interactions of Hamiltonian. Therefore, the direct observable or the value of each peak will be $\propto v_{\theta}/N2^{N-1}$. For example, the direct observation  for $N=4$ will be 32 times as little as that for $N=1$ without interaction. As the size of interacting system increases, it requires higher measurement accuracy. To enhance it, we employed the decoupling detection that cancels the factor of $2^{N-1}$ induced by interactions. Moreover, we swaped all other nuclei to an observable nucleus, sum all experimental decoupling spectra, and measured the combined signal once, which avoids the error caused by multiple readout. Therefore, the final measurement values ($\propto v_{\theta}$) have nothing to do with $N$. The methods utilized will be still valid when extended to larger quantum systems.

\begin{figure}
  \includegraphics[width=8.5cm]{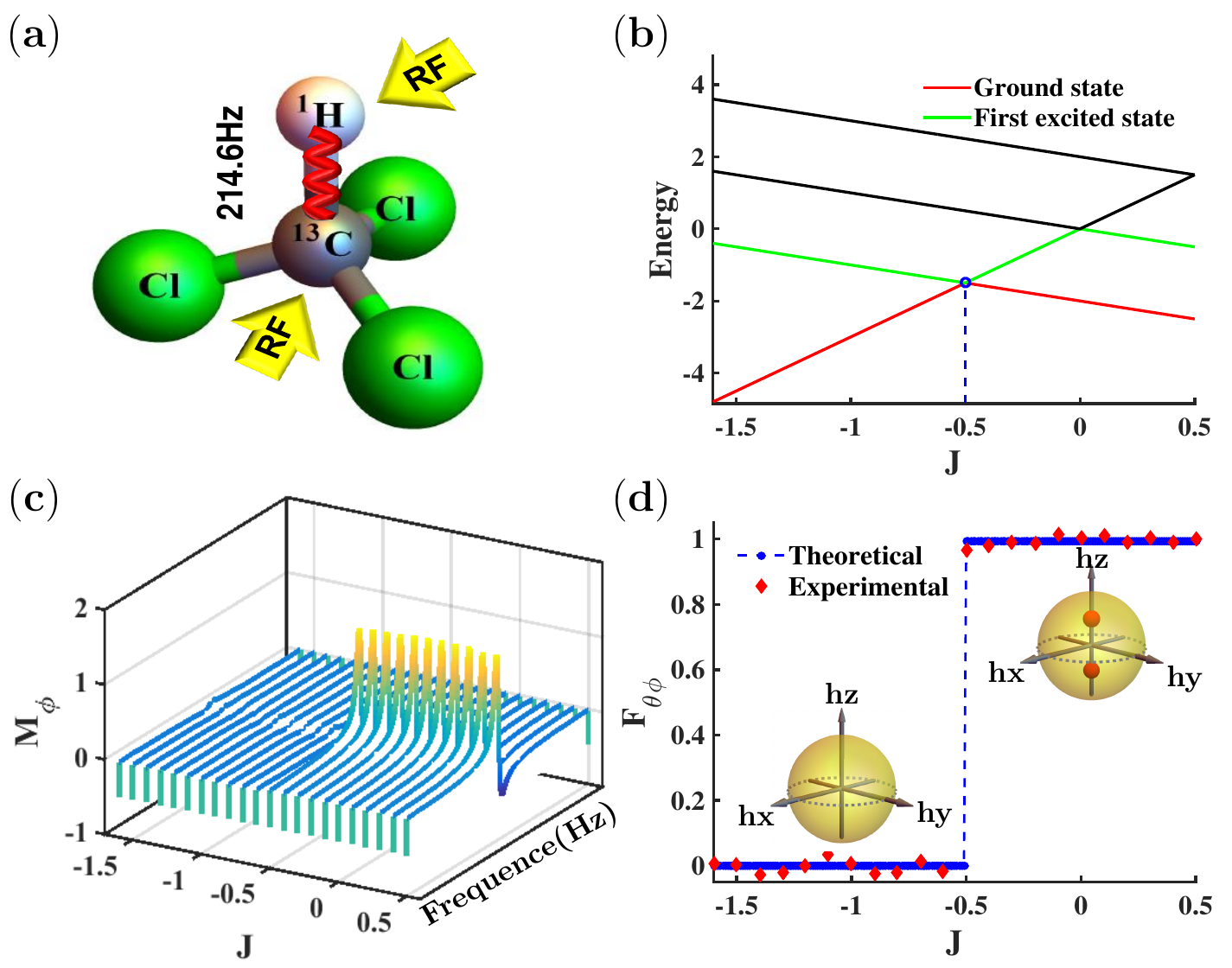}\\
  \caption{(Color online) (a) The molecular structure of Chloroform. The RF pulses act on $^{13}$C and$^{1}$H nuclei independently to to fulfill the desired control tasks. The coupling constant between two nuclei is $J_{\text{CH}}=214.6$ Hz. (b) The energy-level diagram of $N=2$ in Heisenberg spin chain. (c) The sum experimental $^{13}\text{C}$ spectra obtained by decoupling the other spin $^{1}\text{H}$ and swapping $^{1}\text{H}$ to the observed nucleus $^{13}\text{C}$. The longitudinal axis is the integration of the resonant peak of the experimental spectra, which stands for the total magnetization along $y$ direction. (d) The Berry curvature as a function of interaction strength $J$. The blue circles and red diamonds represent the theoretical and experimental values, respectively. $\mathcal{C}h_1 (=2\mathcal{F}_{\phi\theta})$ count the degeneracies (small and red spheres) emerging in $\vec{h}$ parameter space (big and yellow spheres). The experimental average values of different quantized plateaus are $\mathcal{F}_{\theta\phi}=-0.0034\pm0.019$ and $\mathcal{F}_{\theta\phi}=0.99\pm0.015$. }\label{fig:3}
\end{figure}

The experiments were carried out on a Bruker Advance III 400 MHz ($9.4$ T) spectrometer at temperature $303$ K . We first present two-spin experiment using the sample of the $^{13}\text{C}$-labeled chloroform, whose molecular structure is illustrated in Fig. \ref{fig:3}(a). The coupling interaction between $^{13}$C and $^{1}$H nuclei is $\hat{\mathcal{H}}_{zz}=\frac{\pi}{2}J_{\text{CH}}\hat{\sigma}_{\text{C}}^z\hat{\sigma}_{\text{H}}^z$, where $J_{\text{CH}}=214.6$Hz. The Heisenberg spin model of $N=2$ is simulated very well in the short period by using the pulse sequence in Fig. \ref{fig:2}. Its simulated fidelity can be achieved over $0.99$ even if that all pulses  are considered the random errors in the range of 5 degree. The quasiadiabatic evolution path was approximated by $n=300$ discrete steps with reliable accuracy. However, after the whole loop the experimental error will accumulate a lot even if the pulse errors are very small. To overcome this, we packed the loop sequence into one shaped pulse calculated by the gradient ascent pulse engineering (GRAPE) method \cite{Glaser2005}, with the pulse length of 8 ms. The initial ground state at the north pole also was prepared by a GRAPE pulse with pulse length of 5 ms, from the pseudo-pure state (PPS): $\hat{\rho}_{00}=\frac{1-\epsilon}{4}\mathbf{I}+\epsilon| 00 \rangle \langle 00 |$, with $\mathbf{I}$ representing the $4\times4$ identity operator and $\epsilon \approx10^{-5}$ the polarization. PPS $\hat{\rho}_{00}$ was prepared using line-selective approach \cite{Peng2001}, by which the signal strength is larger than that by the spatial average method \cite{Cory1997}. The GRAPE pulses had theoretical fidelities over $99\%$, and were designed to be robust against the inhomogeneity of RF pulses. The sum experimental decoupling spectra for measuring total magnetization are illustrated in Fig. \ref{fig:3}(c). According to the linear response theory, we can obtain the results of $\mathcal{F}_{\phi\theta}$ and further $\mathcal{C}h_1$, as shown in Fig. \ref{fig:3}(d). Figure \ref{fig:3}(b) depicted the energy-level diagram of $N=2$ in Heisenberg spin chain. The energy-level crossing between the ground state and first excited state exactly corresponds to the jumping point of $\mathcal{F}_{\phi\theta}$. It can be seen that the quantized plateaus characterized the interaction-induced topological transition. The first Chern number reveals the number of energy degeneracies emerging in closed manifold of $\vec{h}$ parameter space.

\begin{figure}
  \includegraphics[width=8.5cm]{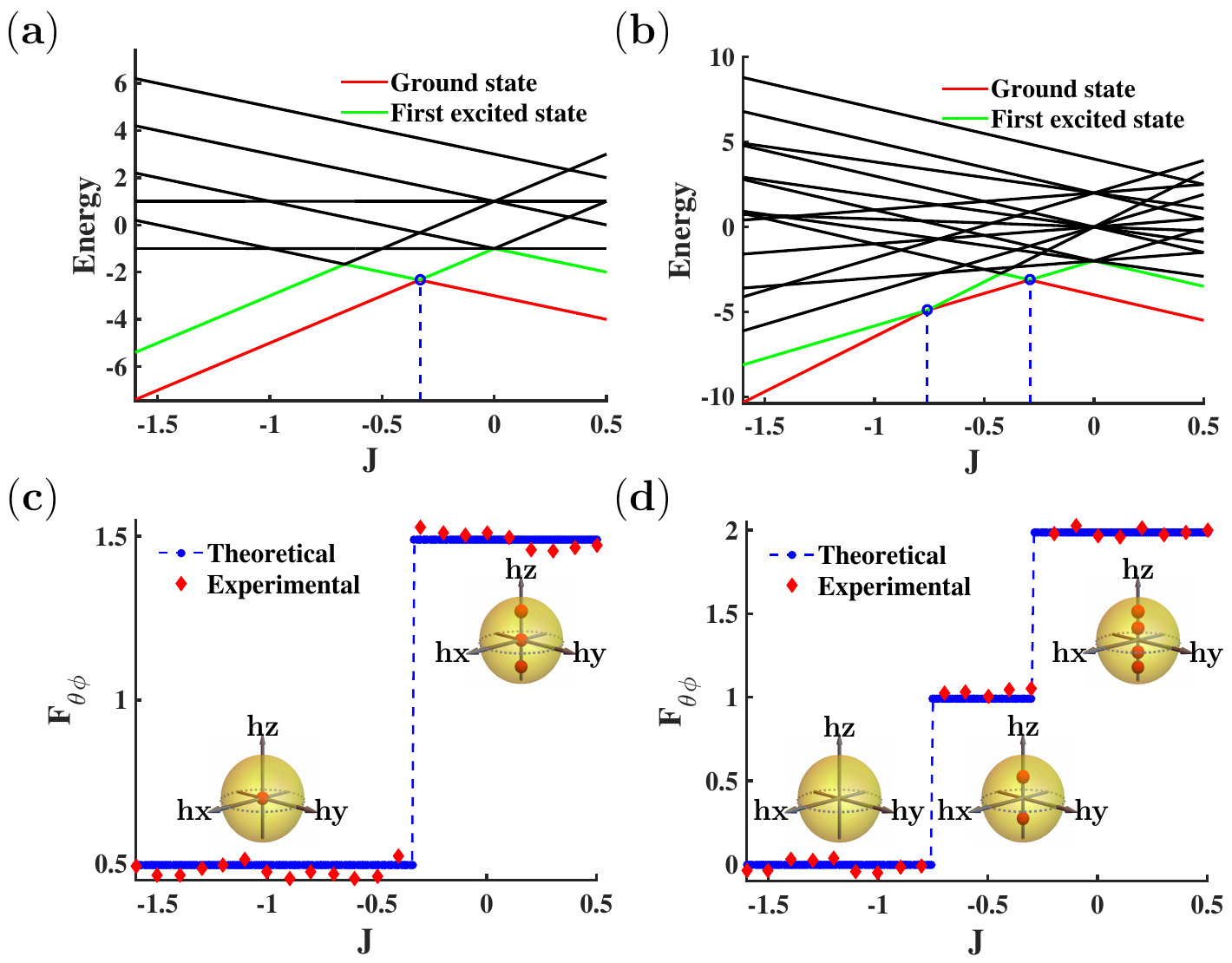}\\
  \caption{(Color online) (a)(b) The energy-level diagrams of Heisenberg spin model for $N=3$ and $N=4$, respectively. (c)(d) The Berry curvatures as a function of interaction strength $J$ in the three- and four-qubit experiments respectively. The blue circles and red diamonds represent the theoretical and experimental values, respectively. The average values of different plateaus are $\mathcal{F}_{\theta\phi}=0.48\pm0.029$ and $\mathcal{F}_{\theta\phi}=1.49\pm0.027$ in (c), and $\mathcal{F}_{\theta\phi}=-0.011\pm0.034$, $\mathcal{F}_{\theta\phi}=1.033\pm0.032$, and $\mathcal{F}_{\theta\phi}=1.99\pm0.024$ in (d), respectively.}\label{fig:4}
\end{figure}

We now turn to three- and four-spin experiments performed on the samples of diethyl-fluoromalonate and iodotrifluoroethylene (see Appendix A). Using the same methods in 2-spin experiment, we measured the total magnetization obtained by integrating the sum experimental decoupling spectra (see Appendix B) and extracted the Berry curvatures shown in Figs. \ref{fig:4}(c) and \ref{fig:4}(d), respectively for $N=3$ and $N=4$. Note that the plateaus in 3-qubit experiment start with nonzero Berry curvature, which is different from the even-spin results. It reflects that there are different degeneracies of ground states in odd-spin and even-spin antiferromagnetic Heisenberg chains \cite{Politi2009,Oh2010}. Moreover, the geometric structure of Hamiltonian can be visualized from the experimental results. That means, without the calculation of Hamiltonian, one can foresee where the level crossings between the ground state and first excited state will happen, as illustrated in Figs. \ref{fig:4}(a) and \ref{fig:4}(b), and how many degeneracies there are inside the closed manifold in $\vec{h}$ parameter space. The first Chern number can be used as nontrivial order parameter to characterize different topological phases and their topological transitions.

These experimental results are in good agreement with theoretical expectations. The relatively minor deviations can be attributed mostly to the imperfections of the preparation of PPS $\hat{\rho_{00}}$ and the spectral integrals. We calculated the standard deviations of experiment and simulation via $\sigma=\sqrt{\sum_{i=1}^M(x_{\text{Exp/Sim}}^i-x_{\text{Th}}^i)^2/M}$. The results are listed in Tab. \ref{tab:1}. The readout error can be estimated by $\sigma_{\text{Read}}=\sigma_{\text{Exp}}-\sigma_{\text{Sim}}^{\text{Tol}}$, which mainly came from spectral integrals. From the two columns of $\sigma_{\text{Exp}}$ and $\sigma_{\text{Sim}}^{\text{Tol}}$, it shows that the controllability became worse as the number of qubits increases. We numerically simulated the errors caused separately by the PPS $\hat{\rho}_0$, ground state preparations and quasiadiabatic evolution, respectively. From the middle results in Tab. \ref{tab:1}, we find that the imperfection of PPS $\hat{\rho}_0$ preparation plays the leading role in the total simulation error. Therefore, it is necessary to prepare high fidelity PPS $\hat{\rho}_0$ in our experiments.

\begin{table}[!bp]
\caption{The standard deviations of experiment, simulation and readout. The total simulation includes PPS $\hat{\rho}_0$, ground state (GS) preparations and quasiadiabatic evolution.}
\label{tab:1}
\centering
\begin{tabular}{|c||c|cccc|c|}
\hline
\hline
\multirow{2}{*}{Qubit} & \multirow{2}{*}{$\sigma_{\text{Exp}}$} & \multirow{2}{*}{$\sigma_{\text{Sim}}^{\text{Tol}}$} &\multirow{2}{*}{$\sigma_{\text{Sim}}^{\text{PPS}}$} &\multirow{2}{*}{$\sigma_{\text{Sim}}^{\text{GS}}$} &\multirow{2}{*}{$\sigma_{\text{Sim}}^{\text{Evol}}$} & \multirow{2}{*}{$\sigma_{\text{Read}}$} \\
&&&&&& \\
\hline
   2 & 0.0171 & 0.0052 & 0.0049 & 0.0046 & 0.0007 & 0.0119 \\ \hline
   3 & 0.0283 & 0.0173 & 0.0159 & 0.0049 & 0.0020 & 0.0110 \\ \hline
   4 & 0.0368 & 0.0243 & 0.0235 & 0.0098 & 0.0070 & 0.0125 \\
  \hline
\end{tabular}
\end{table}

In conclusion, we realized one-dimensional Heisenberg spin model using the interacting nuclear spins and observed the interaction-induced topological transitions in NMR systems. The topological properties of the ground states were analyzed by measuring the Berry curvature and hence the first Chern number. The experimental method utilized for measuring Berry curvature can be used in a variety of generic quantum systems. From the resulting Berry curvature or first Chern number, one can get the geometric information of Hamiltonian about the degeneracies. For instance, the different degeneracies of ground states in odd-spin and even-spin antiferromagnetic Heisenberg chains were observed in Figs. \ref{fig:4} (c) and (d). The quantized plateaus can be applied for precise measurement of the parameter of Hamiltonian. Compared to other platforms such as superconducting systems, NMR systems have the notably advantage in controllability and measurement accuracy, which will provide a better testing platform to explore topological phenomena in more complex quantum systems with interactions. Actually, it is possible for NMR systems with strong coupling interactions, e.g., in low magnetic field ($<10^{-3}$ T)\cite{Appelt2010}, to observe natural topological phases.

This work is supported by NKBRP(2013CB921800 and 2014CB848700), the National Science Fund for Distinguished Young Scholars (11425523), NSFC(11375167, 11227901 and 91021005), the Strategic Priority Research Program (B) of the CAS(XDB01030400), and RFDP (20113402110044).

\section{Appendix}
\subsection{A. Three- and four-qubit quantum registers}

\begin{figure}[!htp]
  \includegraphics[width=8.5cm]{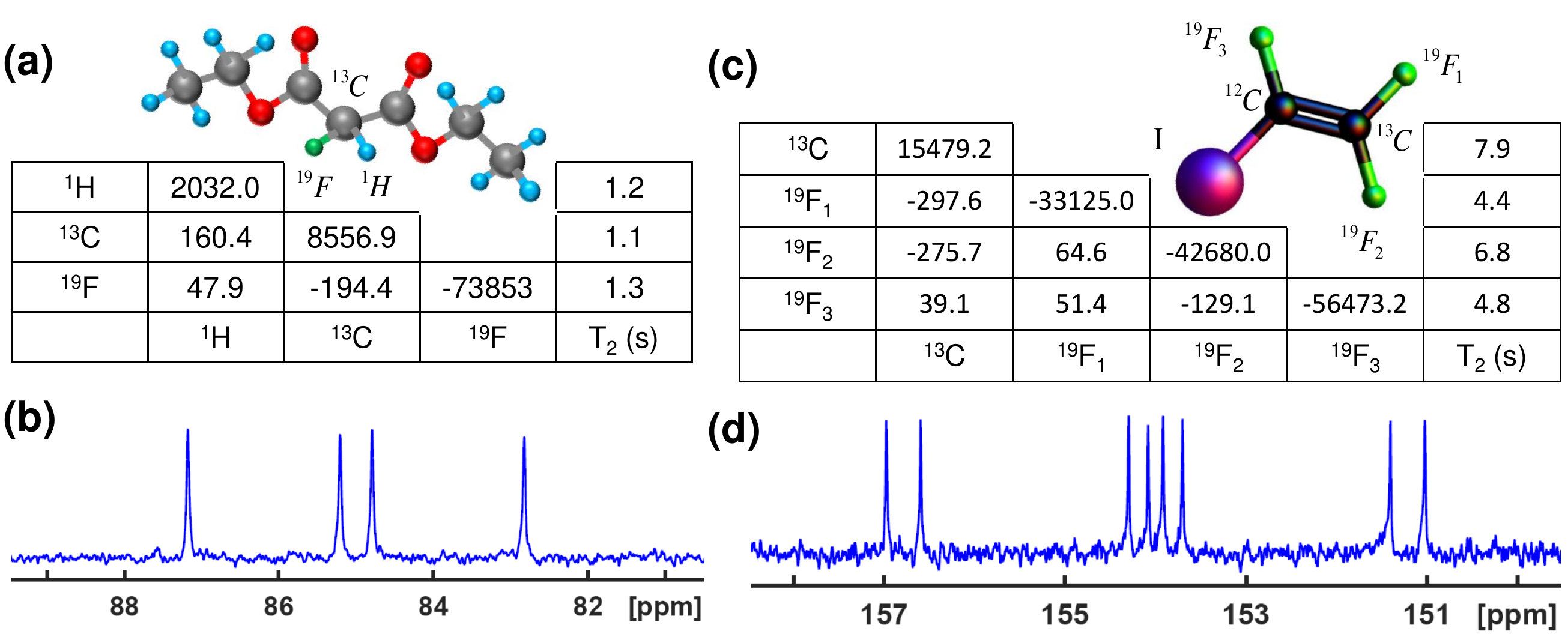}\\
  \caption{(Color online) The molecular structures and parameters of (a) diethyl-fuoromalonate, where three qubits are labeled as $^{1}\text{H},^{13}\text{C and }^{19}\text{F}$ and (c) iodotrifluoroethylene, where four qubits are labeled as $^{13}\text{C},^{19}\text{F}_1,^{19}\text{F}_2 \text{ and } ^{19}\text{F}_3$. The chemical shifts and scalar coupling constants (in Hz) are given as the diagonal and off-diagonal elements in two tables, respectively. The last column shows the transversal relaxation time $T_2$ of each nucleus. Due to the interactions, their corresponding $^{13}\text{C}$ equilibrium spectra of (b) and (d) were splitted into $2^{N-1}$ (i.e.,4 and 8) peaks, respectively. }\label{fig:s1}
\end{figure}

We selected the $^{1}\text{H}, ^{13}\text{C, and } ^{19}\text{F}$ nuclear spins of Diethy-fluoromalonate and one $^{13}\text{C and three} ^{19}\text{F}$ of Iodotrifluoroethylene as the three-qubit and four-qubit registers, whose molecular structures and relevant parameters are listed in Fig. \ref{fig:s1}(a) and (c), respectively. Because both above samples we used are unlabeled, the molecules with a $^{13}$C nucleus (as the quantum registers) were present at a concentration of about $1\%$. The $^{1}$H and $^{19}$F spectra were dominated by signals from the molecules containing the $^{12}$C isotope. To effectively separate this signal from that of the dominant background, it requires to transfer the state of the $^{1}$H and $^{19}$F qubits to the $^{13}$C qubit by a SWAP gate and read the state through the $^{13}$C spectrum. Figure \ref{fig:s1}(b) and (d) are their corresponding $^{13}$C equilibrium spectra.

\subsection{B. The pulse sequences of generating isotropic z-coupling interaction and the sum experimental decoupling spectra for $N=3$ and $N=4$}

\begin{figure}[!htp]
  \includegraphics[width=8.5cm]{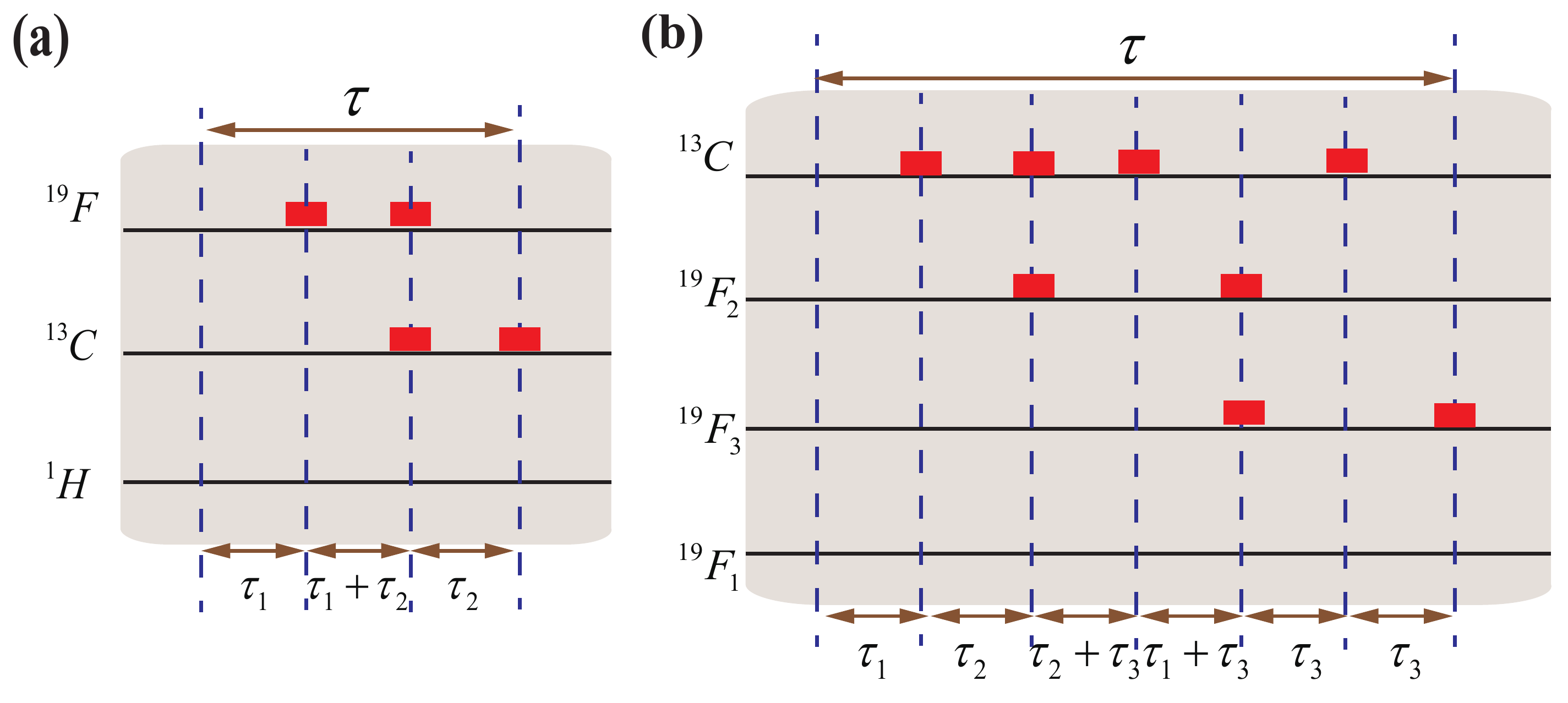}\\
  \caption{(Color online) (a)(b) are the 3-qubit and 4-qubit pulse sequences for effectively creating the isotropic z-coupling interactons, i.e., $\hat{\mathcal{H}}_{zz}=-J\sum_{j=1}^{N-1}\hat{\sigma}_j^z\hat{\sigma}_{j+1}^z$, respectively. The red rectangles represent $\pi$ pulses. In (a), $\tau_1=J_{12}\tau/[2(J_{12}-J_{23})]$ and $\tau_2=-J_{23}\tau/[2(J_{12}-J_{23})]$. In (b), $\tau_1=J_{23}(J_{12}+J_{34})\tau/[4J_{12}(J_{23}-J_{34})]$, $\tau_2=J_{23}(J_{12}-J_{34})\tau/[4J_{12}(J_{23}-J_{34})]$, and $\tau_3=-J_{34}\tau/[4(J_{23}-J_{34})]$}\label{fig:s2}
\end{figure}

\begin{figure}[!htp]
  \includegraphics[width=9cm]{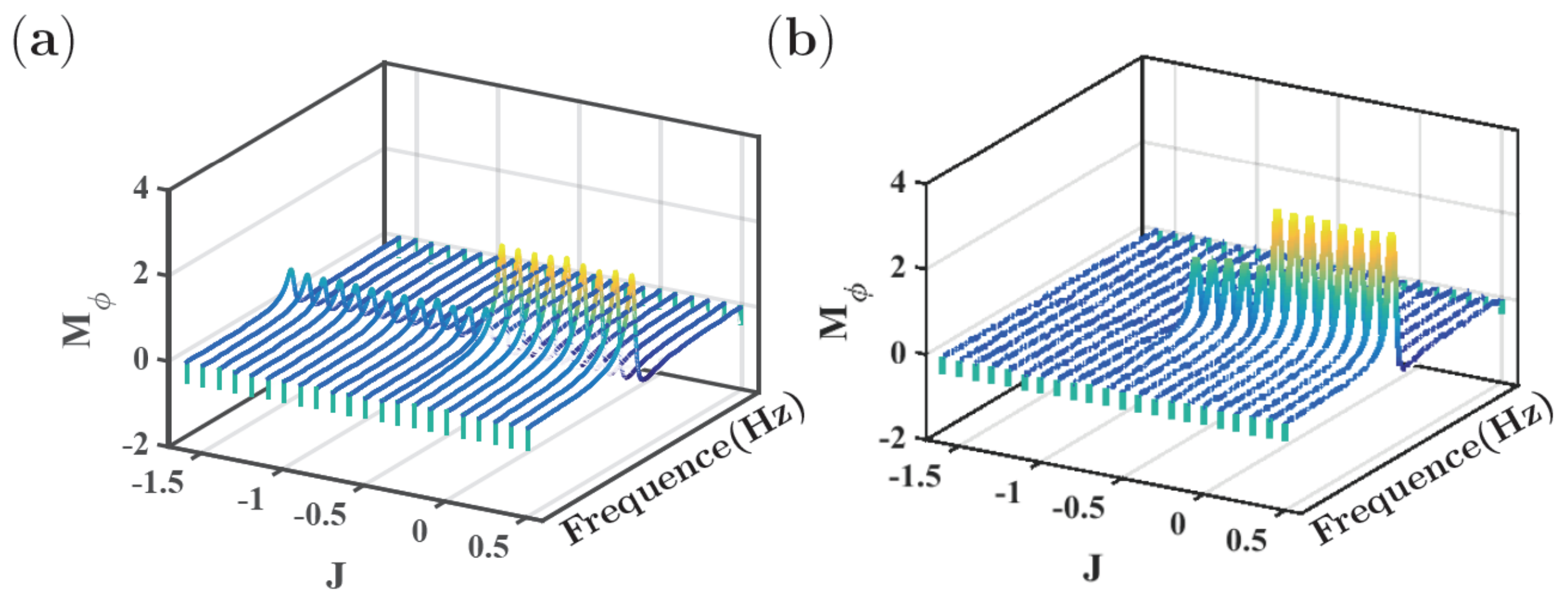}\\
  \caption{(Color online)(a)(b) The sum experimental $^{13}\text{C}$ spectra obtained by decoupling and swapping all other nuclei to the observable nucleus $^{13}\text{C}$, respectively for three- and four-qubit cases.}\label{fig:s3}
\end{figure}

In a short period of $\tau$, we can effectively create the isotropic z-coupling interactions, i.e., $\hat{\mathcal{H}}_{zz}=-J\sum_{j=1}^{N-1}\hat{\sigma}_j^z\hat{\sigma}_{j+1}^z$ only using some refocusing $\pi$ pulses. Figures \ref{fig:s2}(a) and \ref{fig:s2}(b) show two examples of pulse sequences for generating $\hat{\mathcal{H}}_{zz}$ of $N=3$ and $N=4$, respectively. The sum experimental decoupling spectra are illustrated in Fig. \ref{fig:s3}. The integration of the resonant peak of the experimental spectra stands for the total magnetization along $y$ direction. The experimental results show that there are well precise plateaus that reflect the happening of interaction-induced topological transitions.

\end{document}